\documentclass[conference]{IEEEtran}
\IEEEoverridecommandlockouts
\usepackage{cite}
\usepackage{amsmath,amssymb,amsfonts}
\usepackage{algorithmic}
\usepackage{graphicx}
\usepackage{textcomp}
\usepackage{xcolor}
\def\BibTeX{{\rm B\kern-.05em{\sc i\kern-.025em b}\kern-.08em
    T\kern-.1667em\lower.7ex\hbox{E}\kern-.125emX}}
\newcommand{\nn} {\nonumber}
\usepackage{multicol}
\usepackage{multirow}
\begin{document}

\title{3D Channel Modeling and Characterization for Hypersurface Empowered Indoor Environment at 60 GHz Millimeter-Wave Band}

\author{\begin{tabular}[t]{ccc}
\begin{tabular}[t]{c}
Rashi Mehrotra \\ \textit{Department of Computer Science} \\ \textit{University of Cyprus} \\ Nicosia, Cyprus \\ rashi.mehrotra@fau.de
\end{tabular} & 
\begin{tabular}[t]{c}
Rafay Iqbal Ansari \\ \textit{Department of Computer Science} \\ \textit{University of Cyprus} \\ Nicosia, Cyprus \\ ansari.rafay-iqbal@cs.ucy.ac.cy
\end{tabular} &
\begin{tabular}[t]{c}
Alexandros Pitilakis \\ \textit{Department of Electrical} \\ \textit{and Computer Engineering} \\
\textit{Aristotle University of Thessaloniki}\\Thessaloniki, Greece \\alexpiti@auth.gr
\end{tabular} \\
& & \\
\begin{tabular}[t]{c}
Shuai Nie \\ \textit{School of Electrical} \\ \textit{and Computer Engineering} \\ \textit{Georgia Institute of Technology} \\ Atlanta, USA \\ shuainie@gatech.edu
\end{tabular} & 
\begin{tabular}[t]{c}
Christos Liaskos \\ \textit{Institute of Computer Science}  \\
\textit{Foundation of Research} \\ \textit{and Technology Hellas}\\Hereklion, Crete \\cliaskos@ics.forth.gr
\end{tabular} &
\begin{tabular}[t]{c}
Nikolaos V. Kantartzis \\ \textit{Department of Electrical} \\ \textit{and Computer Engineering} \\
\textit{Aristotle University of Thessaloniki}\\Thessaloniki, Greece \\kant@auth.gr
\end{tabular}
\end{tabular} \\
\begin{tabular}[t]{c}
\\
Andreas Pitsillides \\ \textit{Department of Computer Science} \\ \textit{University of Cyprus} \\ Nicosia, Cyprus \\andreas.pitsillides@ucy.ac.cy
\end{tabular}
}

\maketitle 
\begin{abstract}
This paper proposes a three-dimensional (3D) communication channel model for an indoor environment considering the effect of the Hypersurface. The Hypersurface is a software controlled intelligent metasurface, which can be used to manipulate electromagnetic waves, as for example for non-specular reflection and full absorption. 
Thus it can control the impinging rays from a transmitter towards a receiver location in both LOS and NLOS paths, e.g. to combat distance and improve wireless connectivity. 
We focus on the 60 GHz mmWave  frequency band due to its increasing significance in 5G/6G networks and evaluate the effect of Hypersurface in an indoor environment in terms of attenuation coefficients related to the Hypersurface reflection and absorption functionalities, using CST simulation, a 3D electromagnetic simulator of high frequency components. 
To highlight the benefits of Hypersurface coated walls versus plain walls, we use the derived Hypersurface 3D channel model and a custom 3D ray-tracing simulator for plain walls considering a typical indoor scenario for different Tx-Rx location and separation distances.
\end{abstract}

\section{Introduction}
The need and the demand of today's increasingly wireless society for increased bandwidth, reduced latency, and improved robustness motivated a very ambitious set of technical challenges for the next generation mobile wireless systems \cite{imt_itu,Akyildiz_5G}. To meet these new challenges there is a need for new technical innovations. A number of these demanding challenges will benefit with modeling of the channel environment such that the propagation charactersitics, especially for higher frequencies, i.e millimetre wave (mmWave), and in three-dimensional (3D), can be better characterized\cite{3D_channel_3GPP_VTC_2014}. 

The current standard in 3GPP proposed a 3D stochastic channel model for indoor and outdoor environment\cite{3D_mmwave_TMTT}, considering mmWave frequencies.
Moreover, several research efforts have focused specifically in the 60 GHz channel modeling, using a multi-ray based channel model developed for any fixed transmitter and receiver locations and stationary environment\cite{Han_TWC}. Authors in \cite{Josep_TWC}, have proposed a channel model for the THz band. The authors in \cite{Shuai_VTC} have considered a detailed study of the random movement of objects in the environment, as well as the dual mobility of the transmitter and the receiver. This study is particularly important since at mmWave frequencies small objects act as substantial scatterers. These scatterers lead to unwanted non-line-of sight (NLOS) communication. Many researchers have considered the challenges facing the blockage by obstacles such as buildings and furniture in outdoor and indoor environments respectively.  To overcome these problems, they mostly consider using massive MIMO techniques, smart reflect-arrays, and intelligent surfaces (e.g. HyperSurface) \cite{Christos_Mag,Christos_Wowmom} to establish robust mmWave connections for indoor networks, even for the NLOS/obstructed links.
Noteworthy, the authors in \cite{Com_Mag_Shuai} have analysed the distance problem and compared the effect of ultra-massive MIMO, reflect arrays and intelligent surfaces (HyperSurface).

The present work focuses on the HyperSurface (HSF) paradigm for indoor environment with HSF coated walls, which is fully controlled via software, enabling the optimization of propagation factors between wireless devices \cite{Christos_Mag,Christos_Wowmom}.
The methodology proposed in these studies is to coat objects of EM significance in the indoor environment with a novel-class of software controlled metasurfaces, also known as HSF. The contribution of the HSF is that they can locally and adaptively shape the EM behaviour of the environment to solve the distance and non-line-of-sight (NLOS) problem. 
In our implementation, the HSF tiles are rectangular planar thin metasurfaces, composed of tunable sub-wavelength elements called meta-atoms or unit-cells \cite{Christos_Wowmom}.
These elements contain tunable miniaturized controllers (e.g. switches modeled by RLC lumped loads or, generally, complex impedance values), that change the EM behaviour of each unit-cell, and thus collectively alter the EM behaviour of the entire HSF tile. To achieve this functionality, the controllers are networked on the "back" side of the metasurface and a gateway serves as the connectivity unit to provide inter-element and external control \cite{Christos_Mag}.

For the realization of optimal wireless communication networks in the adapted HSF, it is imperative to develop a unified channel model which accurately characterizes the HSF taking into account its peculiarities. The challenges and requirements to be addressed for the analysis and design adapted to the HSF channels can be summarized as follows:
\begin{itemize}
\item Modeling the multi-ray propagation: The multi-ray propagation is present in many common 
scenarios. A unified multi-ray model for the entire mmWave spectrum needs to be developed, which 
incorporates the accurate characterization of the line-of-sight (LoS) and NLOS reflected paths.  
\item Analyzing the channel characteristics: The channel parameters of the adapted HSF at mmWave band were investigated 
via CST simulation tool. These parameters are influenced by multiple factors 
including the operating frequency, communication distance and the material properties of the 
environment. 
\end{itemize}


In this paper, we propose a 3D channel model with intelligent HSF coated walls in the indoor environment, that considers a stationary transmitter (Tx) and a receiver (Rx). For illustration purpose we have assumed far field indoor environment. 
We focus on the 60 GHz mmWave frequency band due to its increasing significance in 5G/6G networks and evaluate the effect of the HSF in an indoor environment in terms of attenuation coefficients related to the HSF reflection and absorption functionalities, using CST, a 3D Electro-Magnetic simulator of high frequency components \cite{cst}.
The impedance parameters of the CST HSF models are tuned using exhaustive search to achieve low values of absorption coefficient, $-26$~dB and $-42$~dB. For reflection coefficient using these impedance parameters, approximately 80 to 93$\%$ of power is reflected using  HSF. 
To highlight the benefits of HSF coated walls versus plain (uncoated) walls, we use the derived HSF 3D channel model and a custom 3D ray-tracing simulator for plain walls considering a typical indoor scenario for different Tx-Rx location and separation distances.
The plain (uncoated) walls are assumed to be normal material, e.g. concrete or brick, with their typical Fresnel coefficients for reflection and absorption.


The paper is structured as follows, section \ref{network} outlines the HSF wireless environment architecture. Section \ref{sec_3D} presents the proposed 3D channel model for indoor environment with the effect of HSF. The HSF parameters are characterized in section \ref{charac_HSF}. Section \ref{simu_res} explains the simulation setup and results. Section \ref{conc} summarizes the conclusion and future directions.

\section{The Hypersurface-based Programmable Wireless Environment Model}
\label{network}
\begin{figure} 
\centering
\includegraphics[width=0.5\textwidth]{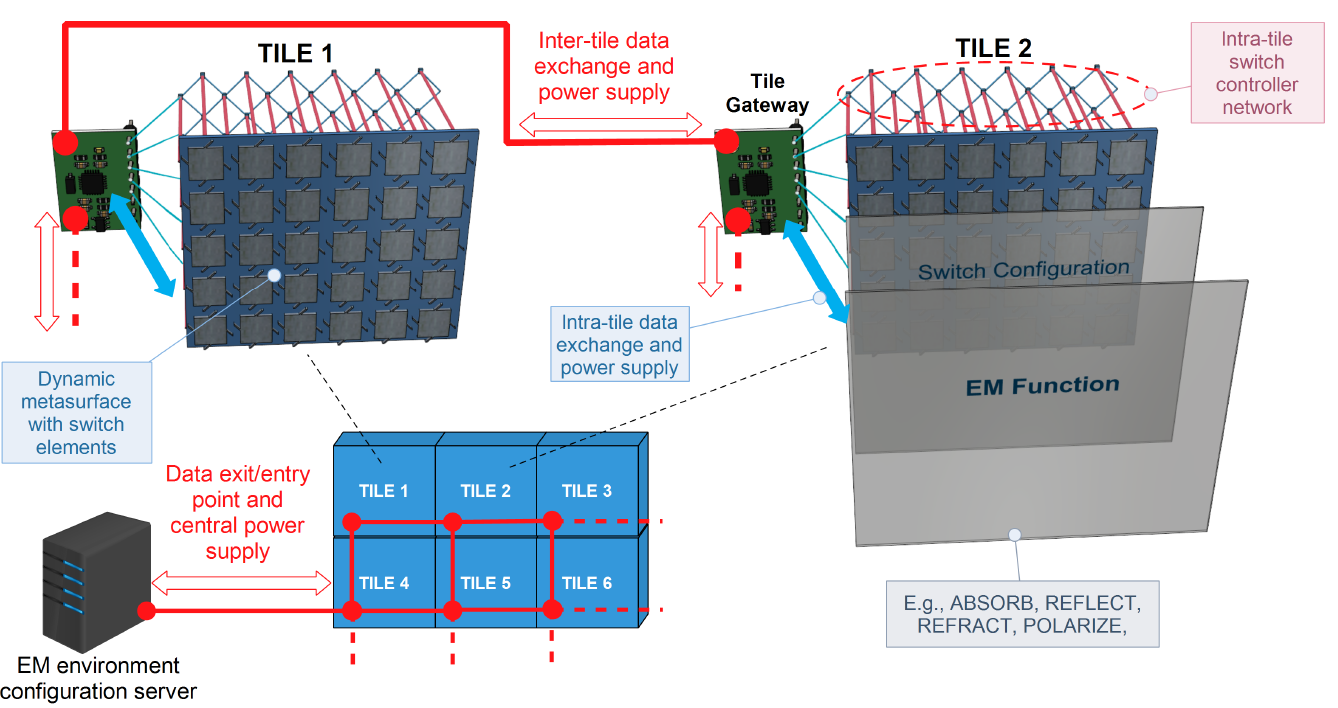}
\vspace{-0.4cm}
\caption{Illustration of the HyperSurface tile architecture and the tile-enabled wireless environment model} 
\label{fig_1}
\vspace{-0.4cm} 
\end{figure}

In this section, we outline the HSF tile hardware components,
the tile intra and inter-networking and the environment control
software \cite{adhoc_christos}. A schematic overview is given in Fig. \ref{fig_1} and is
detailed below. 
As detailed in Fig. \ref{fig_1}, HSF tile has three layers to 
realize all functions required for its use in wireless communication, namely, the metasurface layer, the
intra-tile control layer, and the tile gateway layer \cite{adhoc_christos}. Specifically, the metasurface layer is comprised of the meta-atoms of sub-wavelength size whose configuration is adjustable according to the EM function.
The overall EM behaviour of the meta-atoms is controlled by the combined effect of the tunable miniaturized controllers (switches modeled as impedance values) and the passive metallic patches on their external surface.
The intra-tile control layer is the electronic
hardware component that can control the HSF via software. This layer comprises a network of multiple
controllers, each of which is connected to an active meta-atom element. In particular, the key information, such as
switch configurations among tiles, is exchanged within the network of controllers. The tile gateway layer determines the
communication protocols between the controller network in the intra-tile control layer and the external network. This
architecture of three layers guarantees flexibility and accuracy in the operation workflow of HSFs.

\section{3D Channel Model for Indoor Environment}
\label{sec_3D}

In this section, we consider an indoor environment where all surfaces (e.g. walls, ceilings, objects) are coated with HSF.  The case where surfaces or objects are only partially coated with a HSF are left for future work.
A 3D channel model for indoor environment is proposed that captures the effect of the HSF in the wireless channel. 
The signal propagation in the communication channel includes line-of-sight (LOS) and non-line-of-sight (NLOS) paths, where NLOS are the paths experiencing reflection, refraction and diffraction from the surrounding surfaces, such as walls, furniture etc. In the HSF case, the received signal comprises of the direct path signals as well as any signals reflected by the HSF tiles.
In this paper, we consider reflection and absorption from the HSF covering all surfaces in the indoor environment.

In wireless communication, the received signal as a function of the input signal, the channel response and a random noise\cite{Proakis} can be expressed as 
\begin{align}
r(t)=x(t)*h(t)+n(t)
\end{align}
where $x(t)$ is the transmitted signal, $n(t)$ is the random additive noise and $h(t)$ is the channel response which is given by
\begin{align}
\label{Eq}
h(\tau,\theta,\phi)=\sum\limits_{i=1}^{I}\alpha_{i}\delta(\tau-\tau_i)\delta(\theta-\theta_i)\delta(\phi-\phi_i)
\end{align}
where $\alpha_i$ is the frequency dependent attenuation, $I$ is the number of multi-path components, $\tau$ is the propagation delay, $\tau_i$ is the delay of the $i^{th}$ path. $\theta_i$ and $\phi_i$ are the angles of arrival in elevation and azimuth planes respectively.

The generalized expression of channel response for a given frequency band, including the effect of HSF for the reflection and absorption functions, can be expressed as,
\begin{align}
\label{Eq_main}
h_{rx}(\tau,\theta,\phi)=&\alpha_{LOS}\exp(-j2 \pi f_c\tau_{LOS})\delta(\tau-\tau_{LOS})\\ \nn
+&\sum\limits_{n=1}^{N}\alpha_{ref}^n\alpha_{HSF_{ref}}^{n}(\theta_i,\theta_{ref})
\exp(-j2 \pi f_c\tau_{ref})\\ \nn
\cdot&\delta(\tau-\tau_{ref}^{n})\delta(\theta-\theta_{ref}^{n})\delta(\phi-\phi_{ref}^{n})\\ \nn 
+&\sum\limits_{m=1}^{M}\alpha_{abs}^m\alpha_{HSF_{abs}}^m(\theta_i)\exp(-j2 \pi f_c\tau_{abs})\\ \nn
\cdot&\delta(\tau-\tau_{abs}^{m})
\delta(\theta-\theta_{abs}^{m})\delta(\phi-\phi_{abs}^{m}),
\end{align}
where $N$ and $M$ are the number of tiles set to reflect and absorb respectively. 
In the case of the LOS path, $\alpha_{LOS}$ is the attenuation coefficient of the LOS path, $\tau_{LOS}$ is the delay associated with LOS path and $f_c$ is the operating frequency. For the $N$ tiles set to reflect, $\alpha_{ref}$  is the attenuation coefficient of the reflected component expressed as
$\alpha_{ref}=\left(\frac{c}{4 \pi f_c(d_{tx}+d_{rx})}\right)$  \cite{Han_TWC}, where $c$ is the speed of light,
$d_{tx}$ and $d_{rx}$ are the 3D distance between the transmitter and the HSF tile location and the receiver and the tile location. 
$\tau_{ref}$ is the delay associated with the reflected path (NLOS link) 
which is given by $\tau_{ref}=\frac{d_{tx}+d_{rx}}{c}$, where $\theta_{ref}$ and $\phi_{ref}$ are the angle of reflection in elevation and azimuth planes. In case a tile is set to the absorption mode, $\alpha_{abs}$  is the attenuation coefficient of the absorption component,
$\tau_{abs}$ is the delay associated with this absorption function of the HSF, and $\theta_{abs}$ and $\phi_{abs}$ are the angles due to leakage component in elevation and azimuth planes.
$\alpha_{HSF}$ is the component which depends on the configuration
of the HSF tiles which is quantified in the subsequent
section. In particular, $\alpha_{HSF_{ref}}$ aims to capture the (non-ideal) behaviour of the HSF when set up to reflect at a particular angle of reflection ($\theta_{ref}$) for given angle of incidence ($\theta_i$)  and 
$\alpha_{HSF_{abs}}$ is the absorption component which realistically denotes the leakage due to imperfect absorption at a given $\theta_i$. Note that the most studies only consider the idealized case where $\alpha_{HSF}=1 or 0$. The channel characterization considered in this work strictly applies to the far field (Fraunhofer) region, extending between transmitter and receiver antennas. For 60 GHz radiation, this corresponds to tens of centimeters up to a maximum of a few meters, depending on antenna size. In future work, we will study and further develop the channel/path-loss model including the near field region, which is especially interesting for indoor Hypersurface-empowered network environments.

\section{Characterization of Hypersurface Parameters}
\label{charac_HSF}
In this section, we evaluate and characterize the effect of the HSF in an indoor environment; the parameters we quantify are the $\alpha_{HSF}$ function, the path-length and the reflection (steering) angles. The HSF tiles are composed of unit cells (meta-atoms). For illustration purpose they are designed for the $60$~GHz band with the aim of local and continuous tunability. In order to cover a large spectrum of oblique incidences, operating frequency and both polarizations (normal and parallel to the plane of incidence), we employ an `isotropic' (square) unit-cell made of four square metallic patches with each pair assumed to be connected by a tunable lumped load which models the switch controller (in practice, an integrated circuit, a chip, usually voltage-contolled) see Fig.~\ref{Figure_2_Alex}(a); a sample HSF tile made of a $2\times4$ array of such unit-cells is depicted in Fig.~\ref{Figure_2_Alex}(c). The tunability of the device is implemented by the chips which can be modeled as complex lumped elements (blue arrows in the figure) presenting a series impedance $Z=R+jX$ between two patches, where the reactance ($X$) can in general be either capacitive (negative) or inductive (inductive); in this manner, the surface impedance presented by the HSF can be locally and continuously tuned to accommodate various wavefront-shaping functions \cite{Alex_paper1,Alex_paper2}

The two functions of the HSF targeted in this work are tunable perfect absorption (PA) and tunable anomalous reflection (AR), both applicable to plane wavefronts [\cite{Alex_paper1,Alex_paper2}. For the PA function we require the impinging plane wave frequency, polarization and direction (angle-of-incidence) and, for these parameters, all the unit-cells in the HSF tile are globally tuned to the same $R+jX$ values so that the metasurface can absorb all incoming radiation with minimal reflection or scattering. For the AR function, we additionally require the direction of steering (anomalous, i.e. non-specular, reflection), while the metasurface is configured in supercells, a specific grouping of properly configured unit-cells that is periodically repeated see Fig. \ref{Figure_4_Alex}. Other functions that can be effectuated by the HSF include polarization manipulation and arbitrary wavefront shaping (e.g. focusing of plane waves or collimating spherical wavefronts).

\subsection{Evaluating the $\alpha_{HSF}$ function of Hypersurface}
We characterize the HSF function $\alpha_{HSF}$ given in Eq. (\ref{Eq_main}) for the two tunable functions, perfect absorption (PA) and anomalous reflection (AR), which effectively depend on the configuration of the HSF tiles. $\alpha_{HSF}$ is quantified with the help of the CST EM simulator. To account for LOS and NLOS propagation, it is necessary to characterize the complex-valued coefficient for reflection and absorption of EM waves at 60 GHz mmWave frequency. These coefficients depend on the material properties and geometry of the metasurface, as well as on the frequency and angle of the incident EM wave. We consider the frequency band of $60$~GHz due to its increasing significance for 5G networks. The design and the input parameters of the simulator are the dimensions of the metasurface patches and the optimal resistance and reactance of the chips ($Z=R+jX$) respectively. These values are used to obtain the scattering parameters (S-parameters) corresponding to reflected field amplitudes at all ports/modes of the periodically repeated unit-cells, with the effect of the HSF tiles. The objective is to find the suitable structure dimensions for achieving PA and AR functions, while keeping in view the practicability of the design. We aim for realistic values in our structure parameters thereby providing a platform for easy fabrication. The change in resistance and reactance helps in controlling the surface impedance of the HSF, thereby, we can obtain close to optimal performance (e.g. $-50$~dB reflection for PA and $-0.5$~dB steering in the desired direction for AR) for any given polarization, frequency and direction set, using numerical optimization techniques.

\subsubsection{Absorption}
\begin{figure}
\centering
\includegraphics[width=0.4\textwidth]{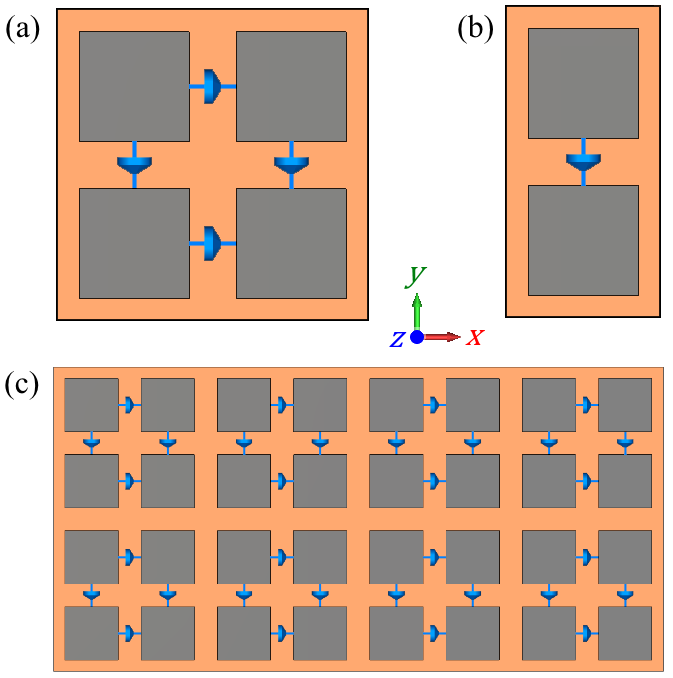}
\vspace*{-0.1cm}
\caption{(a) Square unit cell of 2mm-by-2mm, (b) simplified unit cell, (c) metasurface tile example consisting of $2\times4$ unit cells}
\label{Figure_2_Alex}
\vspace{-0.4cm} 
\end{figure}

\begin{figure}
\centering
\includegraphics[width=0.5\textwidth]{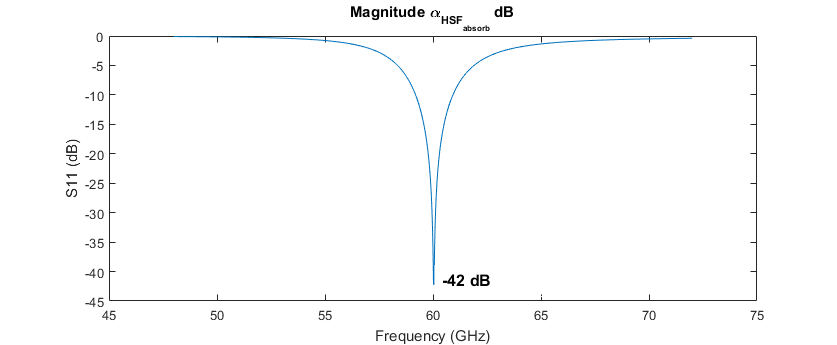}
\vspace*{-0.1cm}
\caption{S11 parameter at 60 GHz}
\label{fig22}
\vspace*{-0.4cm}
\end{figure}

In order to study the absorption function we will assume, for simplicity, that the plane of incidence is the $x0z$ principal plane of the coordinate system of the unit-cell, and the plane wave is TE (normally) polarized, i.e. $\vec{E}=E_y\hat{y}$. Under this assumption, we only need to model half of one unit-cell with only one load, as in Fig.~\ref{Figure_2_Alex}(b). This means that the other three loads (chips) of the square unit-cell are either tuned identically (for the other load parallel to $y$-axis) or open-circuited (for the two loads parallel to $x$-axis). It is straightforward to show how this simplification can be generalized when we have incidence of arbitrarily polarized plane wave impinging from an arbitrary direction. So, the unit-cell considered for absorption at 60 GHz will consist of a pair of copper patches, mounted on a dielectric substrate. The lumped element (the chip) is modeled as a stripe of width $w_{pad}=0.1$~mm of adjustable resistance and capacitance, or inductance. The structure is grounded by a thin metal sheet. The substrate considered for the simulation model has relative permittivity $\epsilon_r=2.2$, loss tangent of $tan\delta=0.0009$ and thickness of 0.127 mm.
The design based on the aforementioned copper patch-based structure is shown in Fig. \ref{Figure_2_Alex}. The square copper patches adopted for absorption have dimensions of 0.75~mm, the gap being 0.08~mm. For analyzing the absorption at normal incidence, we plot the $S_{11}$ parameter (reflection coefficient corresponding to specular direction) to show that absorption is achieved for $60$~GHz mmWave operation. As shown in the Fig. \ref{fig22}, $S_{11}$ = $-42$~dB is attained for normal incidence (elevation angle of incidence is zero). From the simulation model, we identify the magnitude of absorption function ($\alpha_{HSF_{abs}}$ in (\ref{Eq_main})). The value of absorption function for the specific angle of oblique incidence can then be substituted in Eq. (\ref{Eq_main}) for the channel response. Different values of absorption function with respect to different incident angles are shown in table \ref{table_11}.

\begin{table}
\vspace*{-0.1cm}
\caption{Absorption function ($\alpha_{HSF_{abs}}$) for different incident angles}
\begin{center}
\scalebox{0.8}{
\begin{tabular}{|c|c|}
\hline
$\theta_i$ & $\alpha_{HSF_{abs}}$ (dB) \\
\hline
0 & -42 \\
\hline
10&-33\\
\hline
20 & -36 \\
\hline
30& -27 \\
\hline
40& -29\\
\hline
50&-26\\
\hline
60&-28\\
\hline
\end{tabular}
}
\end{center}
\vspace*{-0.4cm}
\label{table_11}
\end{table}

\subsubsection{Anomalous Reflection}
\begin{figure}
\centering
\includegraphics[width=0.4\textwidth]{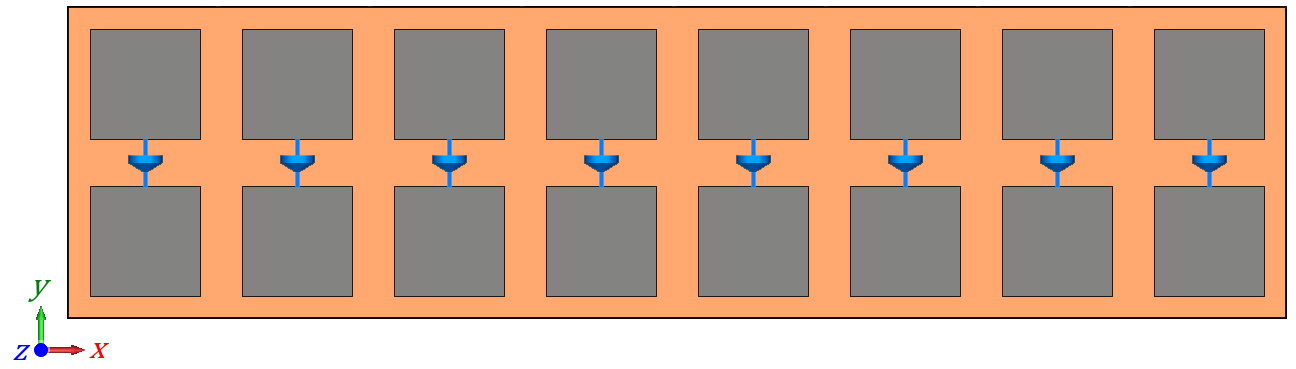}
\vspace*{-0.1cm}
\caption{Super cell structure with 8 unit cells}
\label{Figure_4_Alex}
\vspace*{-0.1cm}
\end{figure}

In a similar manner, we will model steering inside the $x0z$ plane for TE polarized plane waves. For this AR function, we compose a supercell that consists of a stacking of $N_m$ unit cells, where the collective impact of these unit cells leads to a reflected plane wave at an `anomalous' angle, different from specular reflection (e.g. predicted by Snell's law) as shown in Fig \ref{Figure_4_Alex}. The number of unit cells in the supercell varies according to the target steering angle and can be expressed as
\begin{align}
\label{E_6}
sin\theta_{r}-sin\theta_{i}=1/k_0\frac{d\varphi}{dx},
\end{align}
where $\theta_r$ is the angle of reflection (steering) and $k_0$ is the wave-number in vacuum defined as $k_0=2\pi/\lambda$, $\lambda$ being the operating free-space wavelength. Note that Eq.(\ref{E_6}) only prescribes the allowable steering direction(s), not the power steered towards them. To optimize the power steered to the desired direction, we select $\frac{d\varphi}{dx}$ in (\ref{E_6}) by imposing a linear phase profile \cite{Harward_meta} in order to promote the diffraction mode order ($m$), corresponding to the desired steering direction. The direction change can be calculated from the matching of momentum in the $x$-direction, which is given by
\begin{align}
\label{E_7}
\varphi(x)=\varphi_0+m\frac{2 \pi}{D}x,
\end{align}
where $m$ is the diffraction order, $D=N_md_x$, $N_m$ is the number of meta atoms (unit-cells) in the supercell and $d_x$ is the dimension of the meta-atom in x-direction.
Eq. \ref{E_6} gives the relationship of angle of reflection and the number of unit cell in the structure.
Substituting $D=N_md_x$ in (\ref{E_7}) and taking derivative of (\ref{E_7}), we have
\begin{align}
\label{E_8}
\frac{d\varphi}{dx}=\frac{m2\pi}{N_md_x}
\end{align}
On substituting (\ref{E_8}) in (\ref{E_6}) and value of $k_0$, (\ref{E_6}) can be expressed as
\begin{align}
\label{E_9}
sin\theta_{r}-sin\theta_i=\frac{m\lambda}{N_md_x}.
\end{align}
Different values of $N_m$ are obtained from (\ref{E_9}) for particular angle of incidence and reflection (steering) to construct the structure required for anomalous reflection. These values are shown in Table \ref{table2}. The aim of this simulation model is to find $\alpha_{HSF_{ref}}$ by observing the S-parameters. The results of the simulation model will provide us with the S-parameter corresponding to the reflection coefficient in the first diffraction order ($m=+1$), that corresponds to $\alpha_{HSF_{ref}}$.
Furthermore, some losses due to absorption are also quantified that correspond to $\alpha_{HSF_{abs}}$ as given in (\ref{Eq_main}). The values of $\alpha_{HSF_{ref}}$ are shown in Table \ref{table2}. As shown in Table \ref{table2}, over 80$\%$ of reflected power with different values of $N_m$, even for large angle deviations from specular reflection can be expected (note with ‘optimal’ tuning of the impedance values, a higher $\%$ of reflected power can be expected).

\begin{table}
\vspace*{-0.1cm}
\caption{Reflection function $\alpha_{HSF_{ref}}$ for different reflection angles for number of unit cell $N_m$}
\begin{center}
\begin{tabular}{|c|c|c|c|c|}
\hline
$\theta_i$ & $\theta_r$ & $N_m$ & $\alpha_{HSF_{ref}}$(dB) & $\%$ Reflected power  \\ \hline
\hline
\multirow{ 5}{*}{15} & 40 & 13 & -0.521 & 88.70\\
& 50 & 10 & -0.244 & 90.43\\
& 60 & 8 & -0.437 & 90.43\\
& 70 & 7 & -0.768 & 83.79\\
& 80 & 6 & -0.363 & 91.98\\
\hline
\multirow{ 5}{*}{20} & 40 & 16 & -0.882 & 81.62\\
& 50 & 11 & -0.445 & 90.26\\
& 60 & 9 & -0.631 & 86.48\\
& 70 & 8 & -0.552 & 88.06\\
& 80 & 8 & -0.552 & 88.06\\
\hline
\multirow{ 4}{*}{25} & 50 & 14 & -0.818 & 82.83\\
& 60 & 11 & -0.822 & 82.76\\
& 70 & 10 & -0.737 & 84.39\\
& 80 & 8 & -0.897 & 81.34 \\
\hline
\end{tabular}
\end{center}
\vspace*{-0.2cm}
\label{table2}
\end{table}

\section{Evaluation Setup and Results}
\label{simu_res}
In this section, we demonstrate indicative gains offered by using HSF, for a typical indoor scenario, of a stationary transmitter and a stationary receiver, as shown in Fig. \ref{fig6} in comparison to a plain (uncoated) surface.
The results of received power are obtained using HSF using 3D channel model in Eq. (\ref{Eq_main}) populated with values of coefficients as given in Table \ref{table2}.
For the plain setup, the ray tracing model used in \cite{Christos_Wowmom} is adopted. It accounts for first, second, third and fourth order of reflection for the scenarios of the different location of the receivers. Here, the order of reflection denotes the number of bouncing points of the reflected rays. The indoor environment of Fig. \ref{fig6} is ported to the 3D ray-tracing simulator.

The indoor environment setup consists of the room which is separated by the two stacked walls (wall 5 and 6 as in Fig. \ref{fig6}) in the middle. All the walls have height of 4 m, length of 15 m and width of $10$~m. The middle wall is of length $12$~m and thickness $0.5$~m. The location of the transmitter (Tx location) is (7.6, 11.4, 2) m from the reference (0, 0, 0) location, shown in Fig. \ref{fig6}. The transmission power is set to 100 dBmW. We have considered a high power to ensure that no propagation paths are disregarded due to its internal, minimum allowed path loss threshold by the 3D ray tracing simulation. 
The floor and the ceiling are treated as plain,
planar surfaces composed of concrete, without HSF
functionality. In the case of the HSF, all the walls are coated with HSF tiles,
which are square-sized with dimensions 1$\times$1 m. Thus, the 3D
space comprises of a total of 222 tiles.
Please note that the considered tiles functionalities include collimation function. Collimation is the reshaping of a diverging wavefront (typically spherically shaped, originating from a point-source) into a plane wavefront, i.e. a bundle of parallel rays. Thus, the
path loss between two tiles is not subject to the $\propto 1/d^2$
rule, $d$ being their distance \cite{adhoc_christos}. This rule is only valid for the
first impact, i.e., from the transmitter to its LOS tiles.


\subsection{Results and Analysis}

\begin{figure}
\centering
\includegraphics[width=0.4\textwidth]{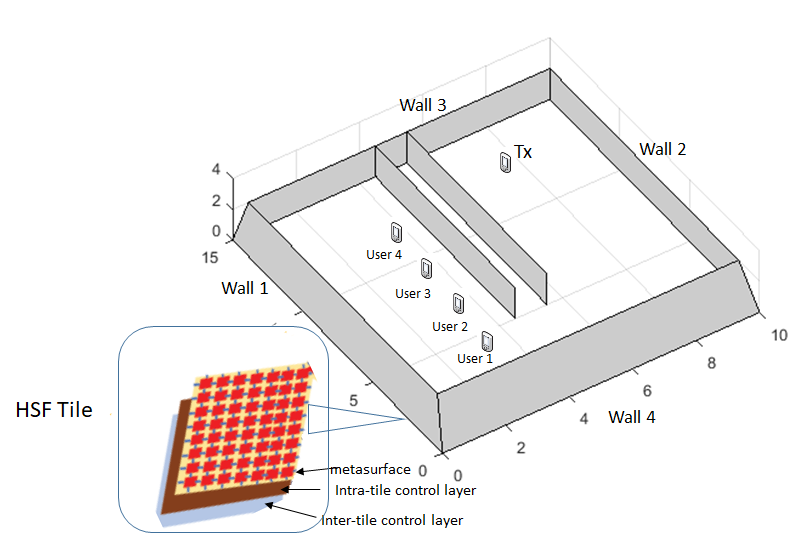}
\vspace*{-0.4cm}
\caption{Simulation Environment with 4 NLOS users} 
\label{fig6}
\vspace*{-0.4cm} 
\end{figure}

In this section, we have demonstrated the analytical model using HSF case with the plain setup using 3D ray tracing simulator showing the indicative gains of HSF coated walls. 
We have considered the NLOS receivers (as in Fig. \ref{fig6}). Four NLOS receiver locations are evaluated.  
The results for received power HSF case considering the value of coefficients from Table \ref{table2} (with respect to the angle of reflection) and the plain (uncoated) cases are tabulated in Table \ref{table5} which shows the values of received power for different values of angle of incidence and reflection for different NLOS receiver locations for the three cases. 
The received power obtained with the HSF case gives a considerable improvement as compared to the walls with no HSF as shown in the last column of Table \ref{table5}. 

\begin{table}[b]
\centering
\small\addtolength{\tabcolsep}{-5pt}
\caption{Comparison of Received Power in $dBmW$ for walls With Ideal and Non-ideal Hypersurface and without Hypersurface}
\vspace*{-0.4cm}
\begin{center}
\scalebox{0.9}{
\begin{tabular}{|c|c|c|c|c|}
\hline
NLOS $R_x$ (m) &  Tile Locations & Plain Setup (dBmW) & Non-Ideal  & $\%$ gain\\ 
&  & & HSF (dBmW)&\\ 
\hline 
(1.15, 0.6, 1.5) & (10, 3.5, 0.5) & 7.23 &  16.411 & 123\\
& (4.5 ,0, 0.5) & & &\\ \hline  
 (1.15, 3.1, 1.5) &  (10, 7.5, 1.5) & 8.24 &  20.391 & 147\\ 
& (3.5, 0, 0.5) & & &\\ \hline
 (1.15, 5.6, 1.5) & (10, 5.5, 1.5) & 7.78 & 17.841 & 129\\ 
 & (4.5, 0, 1.5)& & &\\ \hline 
 (1.15, 8.1, 1.5) & (10, 7.5, 0.5)& 14.91 & 15.159 & 1.67\\ 
&  (5.5, 0, 0.5) & & &\\ \hline	
\end{tabular}
}
\end{center}
\vspace*{-0.4cm}
\label{table5}
\end{table}



%

\section{Conclusion}
\label{conc}
In this paper we develop a 3D channel model that incorporates HSF, computer-controlled active metasurfaces. The design parameters required for the channel model with HSF coated walls are the coefficients for the absorption and reflection functions of the HSF, which are obtained through CST 3D EM simulation of the HSF. 
We have demonstrated the indicative gains of HSF coated walls versus plain (uncoated) walls using the  results obtained using the 3D channel model with the simulation results obtained by a 3D ray-tracing simulator of the plain walls. We show significant improvement in received power for NLOS receivers for different Tx-Rx separation distance with HSF coated walls as compared to walls without HSF in an indoor environment. The proposed 3D channel model can be expanded in the future for partially coated walls and objects, as well as multiple non-stationary users for performance evaluation of different network environments.The expanded model can be used for designing HSF systems as well as for the performance evaluation at $60$~GHz mmWave band.
\vspace*{-0.1cm}

\section*{Acknowledgment}

This work was supported by European Union's Horizon 2020 Future Emerging Technologies call (FETOPEN-RIA) under grant agreement no. 736876 (project VISORSURF).

\bibliographystyle{IEEEtran}
\bibliography{ref}
\end{document}